# Enhancing the Open-Circuit Voltage of Perovskite Solar Cells by Embedding Molecular Dipoles within their Hole-Blocking Layer

*Julian F. Butscher,[1,2] Sebastian Intorp,[3] Joshua Kress,[1,2] Qingzhi An,[1,2] Yvonne J. Hofstetter,[1,2] Nikolai Hippchen,[3] Fabian Paulus,[1,2] Uwe H. F. Bunz,[3]\* Nir Tessler[4] and Yana Vaynzof [1,2]\**

[1] Kirchhoff Institute for Physics and the Centre for Advanced Materials, Heidelberg University, Im Neuenheimer Feld 227, 69120 Heidelberg, Germany

[2] Institute for Advanced Physics and Photonics, Technical University of Dresden, Nöthnitzer Straße 61, 01069 Dresden, Germany

[3] Institute for Organic Chemistry, Im Neuenheimer Feld 270, Heidelberg University, 69120 Heidelberg, Germany.

[4] Sara and Moshe Zisapel Nano-Electronic Center, Department of Electrical Engineering, Technion-Israel Institute of Technology, Haifa 32000, Israel

ABSTRACT

Engineering the energetics of perovskite photovoltaic devices through the deliberate introduction of dipoles to control the built-in potential of the devices offers the opportunity to enhance their performance without the need to modify the active layer itself. In this work, we demonstrate how the incorporation of molecular dipoles into the bathocuproine (BCP) hole-blocking layer of inverted perovskite solar cells improves the device open-circuit voltage ($V_{OC}$) and consequently, its performance. We explore a series of four thiaazulenic derivatives that exhibit increasing dipole moments and demonstrate that these molecules can be introduced into the solution-processed BCP layer to effectively increase the built-in potential within the device, without altering any of the other device layers. As a result the $V_{OC}$ of the devices is enhanced by up to 130 mV with larger dipoles resulting in higher $V_{OC}$s. To investigate the limitations of this approach, we employ numerical device simulations that demonstrate that the highest dipole derivatives used in this work eliminate all limitations on the $V_{OC}$ stemming from the built-in potential of the device.

INTRODUCTION

Lead halide perovskite solar cells have demonstrated remarkable improvements in performance over the last decade, reaching over 25 % power conversion efficiency (PCE) to date.[1] Significant efforts have been dedicated to optimizing the perovskite active layer, with particular emphasis on film composition,[2–5] microstructure[6–8], and reduced defect density.[9,10] Complementing these efforts are improvements in charge extraction layers with a wide range of materials being explored up to now, such as organic polymers,[11–13] small molecules[14–16] and metal oxides.[17–19] It has been shown that both the mobility and the energetic alignment of the charge extraction layers and the perovskite active layer may influence the final device performance,[20] in particular the open-circuit voltage of the device. One of the key challenges in optimizing charge extraction layers originates from the need to preserve the high quality of the perovskite active layer. For instance, it has been shown that the surface energy of the layer, on which the perovskite film is formed, strongly influences its microstructure, electronic structure, and composition.[21,22] As a result, by attempting to change the extraction layer, the active layer may suffer and the overall performance be reduced. Similarly, extraction layers that are deposited from solution on top of the perovskite film should wet the active layer surface and form a homogeneous high-quality film without interfering with the perovskite microstructure or surface structure.[23] To avoid these challenges in device optimization, it is interesting to explore routes to enhance device performance, which do not directly influence neither the active, nor the extraction layers in the device. One such approach focuses on the modification of the extraction layer/electrode interface. While such modifications have been explored in both standard[24–26] and inverted architecture devices,[27–29] the latter received more research attention. In particular, in inverted devices that employ Poly(3,4-ethylenedioxythiophene)-poly(styrenesulfonate) (PEDOT:PSS) as a bottom hole-extraction layer and [6,6]-Phenyl-$C_{61}$-butyric acid methyl ester ($PC_{61}BM$) as an electron extraction layer, early works have shown that there is a need to introduce a hole-blocking layer (HBL) between the $PC_{61}BM$ and the metal electrode in order to achieve a high fill factor.[30] While a range of other materials such as ZnO nanoparticles,[31] TiOx,[30] and the bispyridinium salt FPyBr[32] have been investigated as HBLs in p-i-n perovskite solar cells, bathocuproine BCP remains the most ubiquitously used HBL in such devices[33–37].

Recently, we and others have demonstrated that replacing the BCP layer with a material that introduces a dipole can result in an enhancement of the $V_{OC}$ of the device, provided that the material is still effective in hole-blocking in order to preserve the high fill factor.[38,39] This approach is somewhat limited as only certain materials can satisfy both conditions. Additionally, the relationship between dipole moment of the molecules and the expected increase in $V_{OC}$ remains unexplored.

In this work, we demonstrate that molecular dipoles can be embedded into the BCP layer via simple processing from solution without modifying the overall device structure. We investigate four thiaazulenic derivatives, which were synthesized to exhibit increasing dipole moments. These molecules increase the built-in potential of the device by introducing a dipole at the $PC_{61}BM$/electrode interface. Since the molecules are embedded within BCP, effective hole-blocking is still maintained, securing a high fill factor and short-circuit current. Overall, the $V_{OC}$ of the devices and consequently their efficiency increases gradually with increasing dipole moment. Finally, to investigate the limitations of this method, we performed device simulations that suggest that the maximum possible enhancement of $V_{OC}$ with this method has been achieved, using the thiaazulenic derivatives with the highest dipole moment.

RESULTS AND DISCUSSION

**Synthesis and characterization of thiaazulenic derivatives.**

The four thiaazulenic molecular dipoles (denoted in the following as **TZ-1** to **TZ-4**) were synthesized according to **Scheme 1**. The thiaazulenic core unit is the result of a cyclizing aldol condensation of 3,4-thiophenedicarbaldehydes (**1-3**) with 1,3-substituted acetone derivatives. Subsequent iodination of **4** furnished compound **5**, while **6** was obtained after reversed reaction order from **1** over **2** to **6** due to the reactive thiophene (Th) substituents. **TZ-3** was prepared by first generating 2,5-dimethoxythiophene-3,4-dicarbaldehyde (**3**) followed by condensation with 1,3-difluoroacetone. Stille coupling of the literature known ethyl derivative $7^{40}$ with 2-

(trimethylstannyl)-3,4-ethylenedioxythiophene furnished **TZ-1** in good yields. The remaining thiophene trimers **TZ-2** and **TZ-4** were synthesized in the same manner after Pd-catalyzed coupling with 3-hexyl-5-(tributylstannyl)thiophene. The materials were characterized by standard methods with the results shown in Figures S1-S6 and Tables S2-S5 in the Supplementary Information.

To evaluate the dipole moment of the TZ molecules, molecular electrostatic potential maps were calculated based on their chemical structure (**Figure 1**). The molecules consist of the thiaazulenic core with donor groups (3,4-Ethylenedioxythiophene (EDOT), 3-hexylthiophene, OMe) in 2 and 5 position of the electron rich thiophene and substituents on the electron poor tropone unit that particularly strongly influence the strength of the dipole moment. In **TZ-1**, the ethyl groups as weak donors at the tropone unit and the EDOT groups in the calculated conformation are the reason for a realtively low dipole moment of 1.96 D. By replacing both the EDOT and the ethyl substituent with 3-hexylthiophene and thiophene, respectively (**TZ-2**), the dipole is slightly increased to 3.75 D. The highest dipoles are observed for the **TZ-3** and **TZ-4** derivatives in which strongly electronegative substituents (F, $CO_2Me$) increase the dipole moment to 9.16 D and 10.86 D, respectively. These very high values are in the range of previously reported small molecules with ultrastrong dipole moments.[41]

The TZ materials are soluble in low concentration (e.g. 0.5 mg/ml) in isopropanol making it possible to dissolve them in the BCP solution used for solar cell fabrication. Mixed layers of BCP:TZ were spin-coated on top of the $PC_{61}BM$ electron extraction layer. The incorporation of the molecules was confirmed by X-ray photoemission spectroscopy (XPS) measurements based on the S2p signal from the thiophene units of the TZ molecules with the lowest signal of TZ-3 without additional thiophene-substituents (Figure S7). The surface energies of $PC_{61}BM$/BCP:TZ films were investigated with contact angle measurements (Figure S8). We observe a trend towards lower contact angles for higher dipole moments, which indicates that the TZ molecules are incorporated into the BCP HBL and are at least partly aligned, resulting in the formation of an effective dipole moment.[42] The morphology of the mixed HBL was investigated by atomic-force microscopy (AFM). The AFM micrographs (**Figure 2**) show that the microstructure for all four derivatives of TZ embedded in BCP is similar to that of reference bare BCP layer. This suggests that the TZ molecules do not undergo strong aggregation within the BCP layer and do not form separate domains. Instead, it is likely that the molecules are mixed within the BCP layer in a fairly uniform manner and therefore should not hinder the hole-blocking properties of the BCP layer.

To investigate the influence of the TZ molecules on the energetic alignment with the silver electrode, ultra-violet photoemission spectroscopy (UPS) measurements were performed on $PC_{61}BM/BCP:TZ$ layers. The difference between the work functions (WF) measured on the neat BCP and the mixed BCP:TZ layers is shown in **Figure 3a**. It can be seen that the work function of the BCP layer is gradually increasing in an overall good agreement with the calculated dipole strength of the TZ molecules. This increase in work function will result in an increase in the built-in potential of the device, and should therefore increase its $V_{OC}$. The underlying mechanism is depicted in **Figure 3b**. A dipole introduced by the TZ molecules leads to an upward shift of the vacuum level and thus increases the work function of $PC_{61}BM/HBL$ films. The shift of the vacuum level effectively decreases the work function at the cathode side of the device with respect to the active layer vacuum level, leading to an overall higher built-in potential.

**Photovoltaic performance.**

The effect of an increased built-in potential on the $V_{OC}$ was investigated employing a photovoltaic device architecture as depicted in **Figure 4a** with methylammonium lead triiodide (MAPbI$_3$) as the active layer. The microstructure and crystallinity of the MAPbI$_3$ layer without extraction layers and electrode was characterized by X-ray diffraction (Figures S9a), confirming the high quality of the perovskite active layer. Scanning electron microscopy revealed that the active layer consists of densely packed perovskite grains that protrude through the entire thickness of the film (Figure S9b). The TZ molecules were incorporated within the BCP layer that was deposited on top of the PC$_{61}$BM electron transport layer. **Figure 4c** shows current density–voltage (J–V) curves for representative photovoltaic devices, measured in the dark. A significant shift of the "knee" position of the curves to higher voltages for increasing dipole moment indicates an enhancement of the built-in potential because flat band conditions are matched at higher voltages. To investigate if this trend is maintained under illumination, J-V curves were measured under AM 1.5 G simulated 100 mW cm$^{-2}$ sunlight. The J-V characteristics for optimal devices (meaning here devices of optimal V$_{OC}$) are shown in **Figure 4d** with the corresponding external quantum efficiency curves shown in Figure S10. As expected, the V$_{OC}$ rises with increasing dipole moment to a maximum value of 1.09 V for BCP:TZ-3 and BCP:TZ-4 HBLs which represents an increase of 130 mV when compared to the optimal reference device with only BCP as HBL. We note that although the average work function shift of TZ-3 is only 0.1 eV larger than that of TZ-2 (Figure 3a), the V$_{OC}$ of TZ-3 (1.09 eV) is noticeably higher than that of TZ-2 (1.02 eV). It is possible that this originates from an error associated with UPS measurements, or slight sample-to-sample variation between the photovoltaic devices and samples for UPS characterization. The qualitative trend, however, is maintained, with larger molecular dipoles resulting in higher V$_{OC}$s.

Equally important, the short-circuit current (J$_{SC}$), the fill factor (FF) and the overall shape of the J-V curves is maintained when incorporating the thiaazulenic compounds. This confirms that the hole-blocking properties of BCP are unaffected by the incorporation of TZ molecules since otherwise the FF of the devices would decrease and the devices show an S-shape as previously reported for devices without HBL.[43] This is also confirmed by UPS measurements that demonstrate that the valence band structure of the BCP:TZ layers is comparable to that of neat BCP (Figure S11). The photovoltaic parameters for both reverse and forward scans of the devices are listed in **Table 1**. It is noteworthy that all of the J-V curves show a rather low hysteresis as it is expected for devices with a $PC_{61}BM$ electron extraction layer. An overview of the performance data over a total of 40 devices is shown in **Figure 5**, confirming the reproducibility of this approach. A maximum power conversion efficiency of 18.8 % was achieved for a device with BCP:TZ-4 as HBL yielding a $V_{OC}$ of 1.07 V. The average $V_{OC}$ for devices employing BCP:TZ-3 and BCP:TZ-4 as HBL (**Figure 5a)** is very similar suggesting that a maximum improvement was already achieved with BCP:TZ-3.

**Numerical simulations.**

To confirm our experimental results as well as to probe the limitations of this method we employed numerical device simulations.[44] Such simulations are based on solving numerically the drift-diffusion and Poisson equations in the perovskite solar cell with energy level diagrams as shown in Figure S12, in which the change in WF (ΔWF) at the BCP:TZ layer has been varied from 0 eV (representing the near BCP case) to 0.5 eV in steps of 0.1 eV. Table S1 summarizes the other parameters used for the simulation. In **Figure 6a** all simulated J-V curves for different ΔWFs are shown. Similar to the experimental J-V data, shown in Figure 4d, an increase in $V_{OC}$ is observed with an increasing WF/dipole moment while $J_{SC}$ and FF remain rather constant leading to an overall improvement of the PCE. **Figure 6b** summarizes the increase in $V_{OC}$ - with respect to the reference device - against an increase in ΔWF. The results suggest a trend, in which beyond a certain point a further increase in ΔWF no longer improves the $V_{OC}$. We note that the device simulations suggest a possible increase of only about 80 mV, however, this absolute value depends on the simulation parameters, while the overall observed trend does not. The similarity in the trend of both theory and experiment suggests that the $V_{OC}$ plateau, reached for devices with BCP:TZ-3 and BCP:TZ-4, represents a limit to the maximum obtainable $V_{OC}$ increase obtainable using this method. As a result, further increasing the built-in potential cannot improve the device $V_{OC}$ since other effects such as e.g. bulk and surface recombination become more dominant[45–48]. Nevertheless, an increase of 130 mV in $V_{OC}$ is a significant improvement, which might also be applicable for other perovskite devices [49–51].

It is important to note that the built-in potential in perovskite solar cells does not represent a hard limit on the $V_{OC}$[52]. For instance, blocking layers can recover the loss due to a low built-in potential through electrostatic band bending that farther separates the electrode energies beyond the built-in potential[53]. However, this recovery mechanism does not completely compensate for these losses leaving space for further improvements through band engineering[39,54]. The saturation of the $V_{OC}$ in both experimental (Figure 5a) and simulation (Figure 6b) results suggests that the limitations by built-in potential can be fully eliminated if sufficiently large dipoles are introduced in the device hole-blocking layer.

**Conclusion.**

In this work we investigated the effect of introducing thiaazulenic molecules with a dipole moment into the BCP hole-blocking layer of MAPbI$_3$ solar cells without altering the overall device structure. We demonstrated that the work function of $PC_{61}BM$/HBL can be systematically increased by introducing molecules with increasing dipole moment, leading to an increase of the built-in potential in the final device. Consequently, the $V_{OC}$ improves for higher dipole moments, yielding an impressive increase of up to 130 mV. A maximum $V_{OC}$ of 1.09 V could be realized with this technique, a value compatible with devices employing PTAA[55–57] or $NiO_x$[58–60] as hole transport layers. This finding confirms that the controlled embedding of dipole molecules within the hole-blocking layer can compensate for the lower work function of hole transport layers such as PEDOT:PSS and eliminate the limitations to the $V_{OC}$ that stem from it. Finally, numerical device simulations reveal that the maximum possible improvement of the $V_{OC}$ is achieved with the high dipole moment molecules TZ-3 and TZ-4.

FIGURES

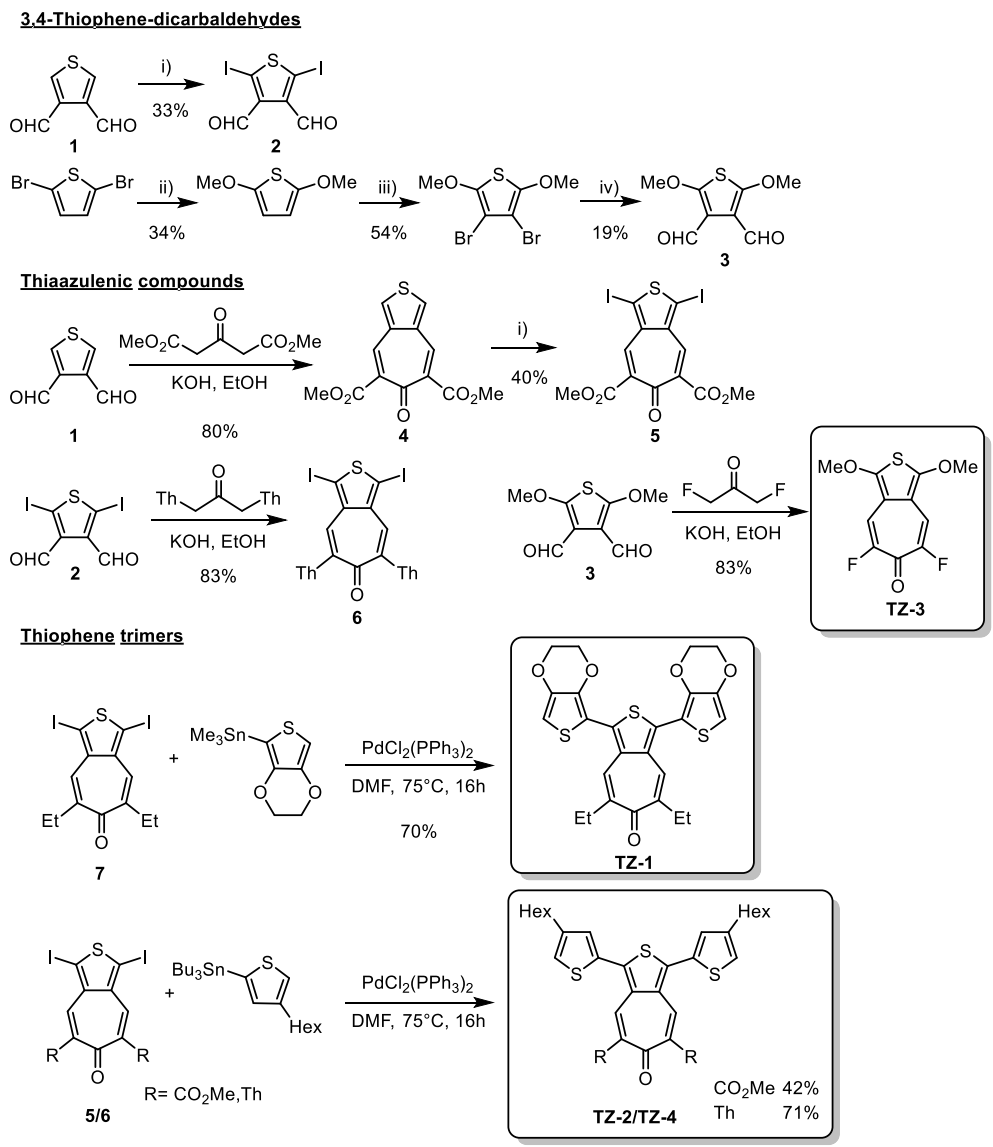

**Scheme 1**. Synthesis of thiaazulenic dipole molecules TZ-1-4. i) I$_2$, PhI(CF$_3$COO)$_2$, DCM, rt; ii) NaOMe (25%), CuCl, HCOOH, MeOH, 120°C; iii) NBS, CCl$_4$, rt; iv) 1.*n*-BuLi, DMF, 2. *t*-BuLi, DMF, 3. HCl, Et$_2$O, −78°- rt.

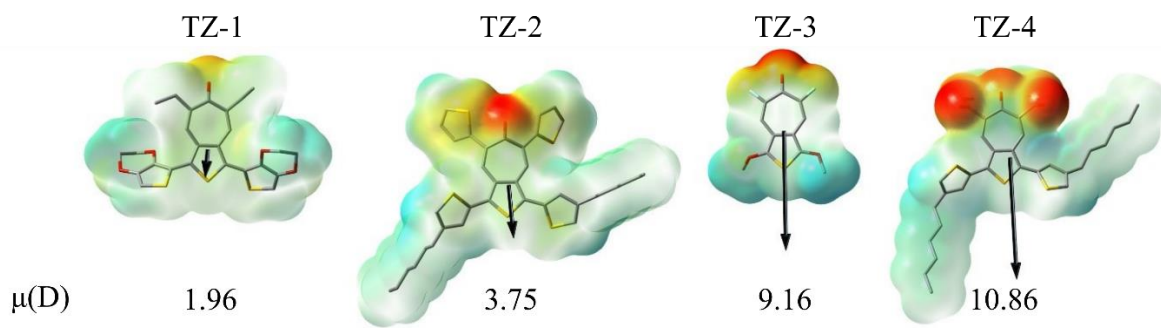

**Figure 1**. Molecular electrostatic potential maps of the four thiaazulenic derivatives with red indicating strongly and blue weakly electronegative parts of the molecules.

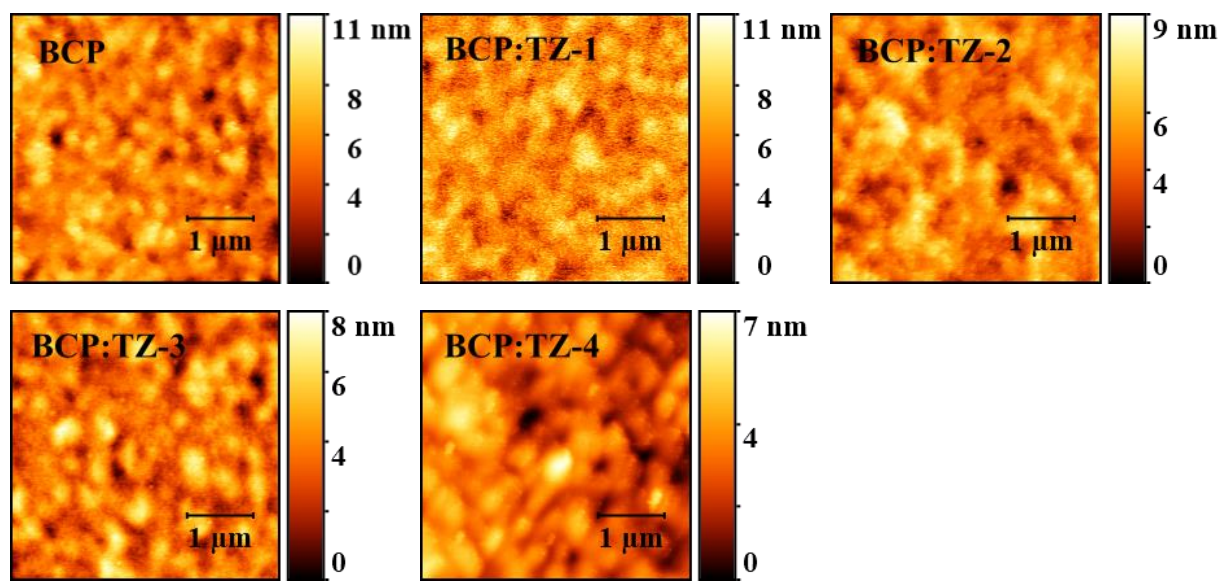

**Figure 2**. Atomic force microscopy images of BCP and BCP:TZ films in ascending order of their dipole moment. All films exhibited a similar RMS roughness value in the range of 1-1.4 nm.

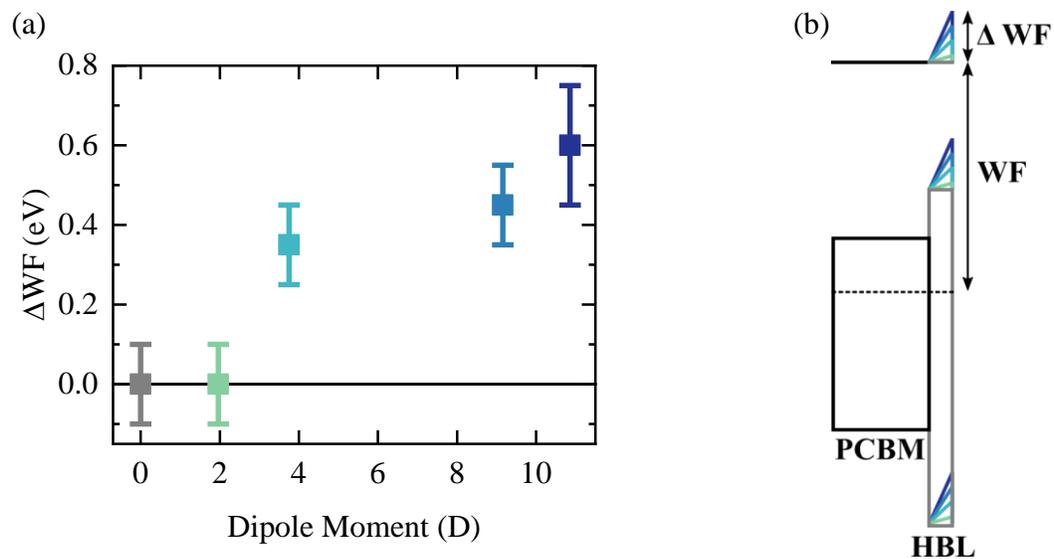

**Figure 3**. (a) Work function evolution of $PC_{61}BM/BCP:TZ$ films with respect to bare $PC_{61}BM/BCP$ films as a function of their dipole moment and (b) schematic illustration of the underlying mechanism that leads to an increase in built-in potential.

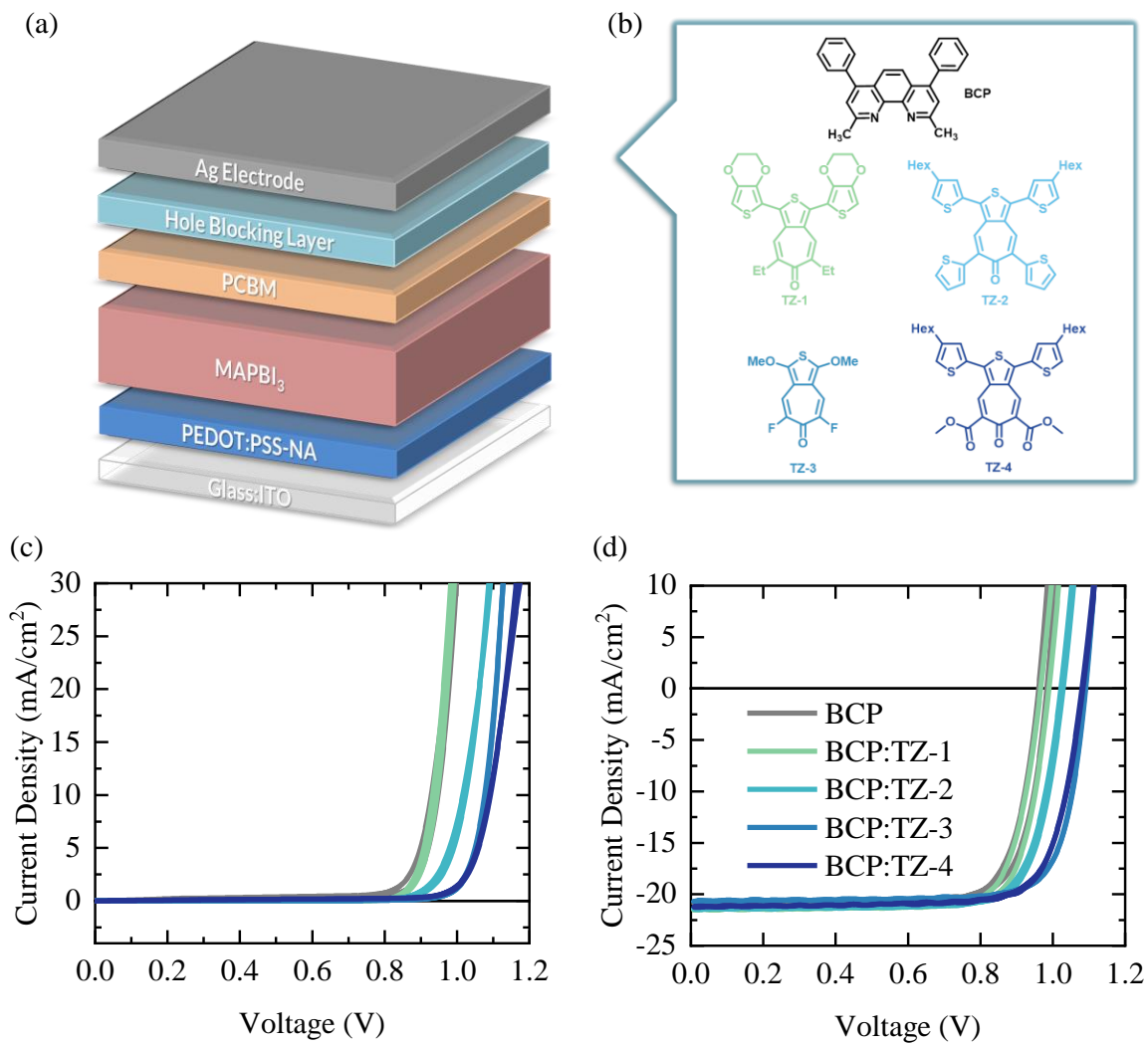

**Figure 4**. Photovoltaic performance of devices with either BCP or BCP:TZ hole-blocking layers. (a) Schematic PV device structure (b) BCP and TZ molecules 1-4 c) J-V curves of representative devices measured in the dark (d) J-V curves of optimal devices measured under AM1.5 illumination.

**Table 1**. Photovoltaic parameters of optimal MAPbI$_3$ devices with either BCP or BCP:TZ HBLs. FS and RS represent scanning direction from J$_{SC}$ to V$_{OC}$ and V$_{OC}$ to J$_{SC}$, respectively.

|  | Dipole moment [D] | V$_{OC}$ FS [V] | J$_{SC}$ FS [mA/cm$^2$] | FF FS [%] | PCE FS [%] | V$_{OC}$ RS [V] | J$_{SC}$ RS [mA/cm$^2$] | FF RS [%] | PCE RS [%] |
|---|---|---|---|---|---|---|---|---|---|
| **BCP** | 0 | 0.96 | -20.90 | 79.09 | 15.89 | 0.98 | -20.90 | 78.97 | 16.24 |
| **BCP:TZ-1** | 1.96 | 0.97 | -21.58 | 78.20 | 16.32 | 0.99 | -21.58 | 79.06 | 16.89 |
| **BCP:TZ-2** | 3.75 | 1.02 | -21.37 | 78.40 | 17.15 | 1.03 | -21.37 | 78.69 | 17.31 |
| **BCP:TZ-3** | 9.16 | 1.09 | -20.77 | 79.33 | 17.96 | 1.09 | -20.77 | 79.25 | 17.97 |
| **BCP:TZ-4** | 10.86 | 1.09 | -21.15 | 77.11 | 17.73 | 1.09 | -21.15 | 76.51 | 17.63 |

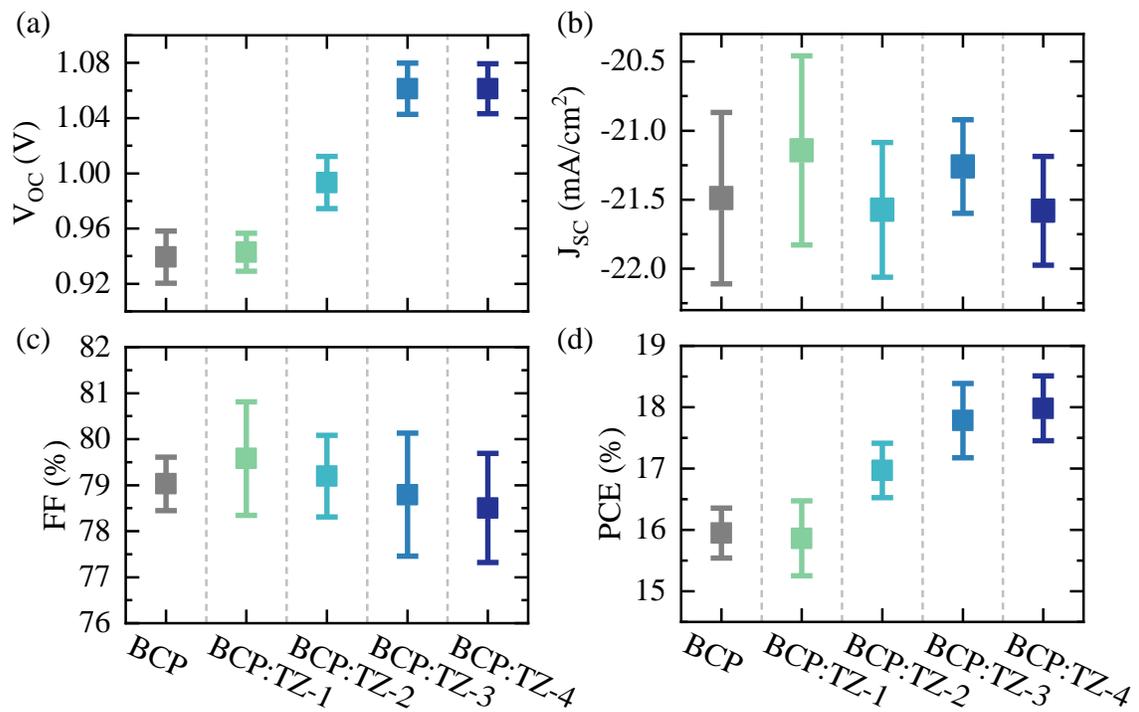

**Figure 5**. (a) V$_{OC}$, (b) J$_{SC}$, (c) FF and (d) PCE of 40 photovoltaic devices with BCP:TZ layers and reference BCP only devices.

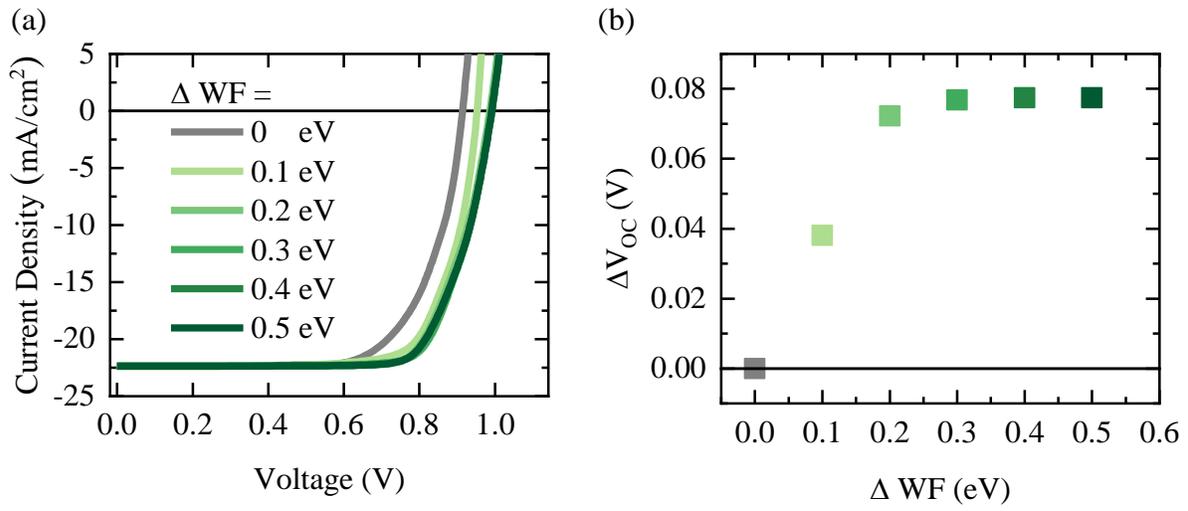

**Figure 6**. (a) Simulated J-V-Curves for different work function shifts and (b) corresponding increase in V$_{OC}$ from the simulation results with respect to no dipole moment.

EXPERIMENTAL METHODS

**Materials:** Methylammonium iodide ($CH_3NH_3I$) was purchased from GreatCell Solar. PEDOT:PSS was purchased from Heraeus Deutschland GmbH&Co and $PC_{61}BM$ (99.5%) was purchased from Solenne BV. All other materials were purchased from Sigma-Aldrich and used as received.

**Photovoltaic Device Fabrication:** Pre-patterned indium tin oxide (ITO) coated glass substrates (PsiOTech Ltd., 15 Ohm/sq) were ultrasonically cleaned with 2 % hellmanex detergent, deionized water, acetone, and isopropanol, followed by 8 min oxygen plasma treatment. PEDOT:PSS prepared based on the previous report[12] was spin coated on the clean substrates with 4000 rpm 30 s and annealed at 150 °C for 15 min. A lead acetate trihydrate $MAPbI_3$ recipe[17] was used for forming the $MAPbI_3$ perovskite layer, in detail the perovskite solution was spin coated at 2000 rpm for 60 s in a dry air filled glovebox (RH < 0.5 %). After blowing 25 s and drying 5 min, the films were annealed at 100 °C for 5 min forming a black and uniform perovskite layer. The prepared samples were transferred to a nitrogen filled glovebox, $PC_{61}BM$ (20 mg/ml dissolved in chlorobenzene) was dynamically spin coated at 2000 rpm for 30 s on the perovskite layer followed by a 10 min annealing at 100 °C. Sequentially either BCP or a blend of BCP and the thiaazulene derivatives at a 1:1 weight ratio (0.5 mg/ml dissolved in isopropanol) was spun on the $PC_{61}BM$, forming a thin layer of around 5 nm. To complete the device, 80 nm silver was deposited via thermal evaporation under high vacuum.

**Photovoltaic Device Characterization:** The current density-voltage (J-V) was measured by a computer controlled Keithley 2450 Source Measure Unit under simulated AM 1.5 sunlight with 100 mW cm$^{-2}$ irradiation (Abet Sun 3000 Class AAA solar simulator). The light intensity was calibrated with a Si reference cell (NIST traceable, VLSI) and corrected by measuring the spectral mismatch between the solar spectrum, the spectral response of the perovskite solar cell and the reference cell. The mismatch factor was calculated to be 10 %. The cells were scanned from forward bias to short circuit and reverse at a rate of 0.25 V s$^{-1}$ by employing a mask to eliminate the overestimate of the photocurrent.

**UV-vis absorption:** The spectra of the thiaazulene molecules were measured in solution using isopropanol as the solvent. The UV-vis spectra were measured at room temperature with a Jasco UV-Vis V670 spectrometer.

**Ultra-violet Photoemission Spectroscopy (UPS):** Ultra-violet photoemission spectroscopy measurements were performed on BCP/PC$_{61}$BM and BCP:TZ/PC$_{61}$BM films to characterize their work function. The samples were transferred to an ultrahigh vacuum (UHV) chamber of the PES system (Thermo Scientific ESCALAB 250Xi) for measurements. UPS measurements were carried out using a double-differentially pumped He discharge lamp (hν = 21.22 eV) with a pass energy of 2 eV and a bias of -10 V.

**Atomic Force Microscopy (AFM):** AFM (Bruker MultiMode) was performed in ScanAnalyst mode in air with silicon tips (Bruker NTESPA) to study the surface morphology of the reference BCP and BCP:TZ films.

**Contact Angle Measurements:** Contact angle measurements were performed using a Ramé-Hart 260 goniometer with an automated dispensing system for water. A water droplet was created and placed on the substrates. The sessile drop was imaged and the contact angle was extracted using the Ramé-Hart DROPimage Advanced software.

**Scanning Electron Microscopy (SEM):** SEM was performed using a Jeol M-7610F FEG-SEM. Perovskite films were mounted on brass SEM holders using conductive carbon tape and liquid silver paste to avoid sample charging. Top View images were recorded at 1.5 kV with the low in chamber secondary electron detector (LEI) to image mainly the topography of the sample. Cross-section images were taken from devices cut in half at 5.0 kV with an 'in-lense' equivalent detector (SEI), collecting a combination of secondary and backscattered electrons.

**X-ray Diffraction:** The MAPbI$_3$ active layer on glass/ITO/PEDOT:PSS-Na were characterised using a Rigaku Smart Lab diffractometer with a rotating copper anode (9 kW, 45 kV, 200 mA) equipped with a 2D HyPix3000 detector (with a 0.2 mm𝑓 collimator) in a coupled theta-2theta scan. Diffraction maps were integrated to obtain a 1D-diffraction profile and background (empty glass/ITO substrate) corrected using the Rigaku 2DP software.

**Device Simulation:** A previously developed model that includes the contributions of charges and ions has been used to simulate the influence of incorporating the TZ derivatives into the BCP HBL.[44] Further details are provided in the Supplementary Material.

ASSOCIATED CONTENT

**Supporting Information**. The following files are available free of charge: UV-vis of thiaazulene derivatives in solution; scanning-electron microscopy image of MAPbI$_3$ perovskite active layer; X-ray diffraction pattern of MAPbI$_3$ perovskite active layer; (Additional photovoltaic device statistics;) Contact angle measurements; device simulation parameters.

AUTHOR INFORMATION

**Corresponding Authors**

*Email: yana.vaynzof@tu-dresden.de

*Email: uwe.bunz@oci.uni-heidelberg.de

**Funding Sources**

This work has received financial support of the Deutsche Forschungsgemeinschaft (DFG) SFB 1249 projects A03 and C04. This project has also received funding from the European Research

Council (ERC) under the European1 Union's Horizon 2020 research and innovation program (ERC Grant Agreement n° 714067, ENERGYMAPS). N.T. acknowledges support by the Israel Science Foundation (grant no. 488/16), the Adelis Foundation for renewable energy research within the framework of the Grand Technion Energy Program (GTEP), and the Technion Ollendorff Minerva Center.
REFERENCES

(1) NREL. Research Cell Efficiency Records https://www.nrel.gov/pv/cell-efficiency.html (accessed Nov 13, 2019).

(2) Saliba, M.; Matsui, T.; Domanski, K.; Seo, J.-Y.; Ummadisingu, A.; Zakeeruddin, S. M.; Correa-Baena, J.-P.; Tress, W. R.; Abate, A.; Hagfeldt, A.; Gratzel, M. Incorporation of Rubidium Cations into Perovskite Solar Cells Improves Photovoltaic Performance. *Science* **2016**, *354* (6309), 206–209.

(3) Saliba, M.; Matsui, T.; Seo, J. Y.; Domanski, K.; Correa-Baena, J. P.; Nazeeruddin, M. K.; Zakeeruddin, S. M.; Tress, W.; Abate, A.; Hagfeldt, A.; Grätzel, M. Cesium-Containing Triple Cation Perovskite Solar Cells: Improved Stability, Reproducibility and High Efficiency. *Energy Environ. Sci.* **2016**, *9* (6), 1989–1997.

(4) Singh, T.; Miyasaka, T. Stabilizing the Efficiency Beyond 20% with a Mixed Cation Perovskite Solar Cell Fabricated in Ambient Air under Controlled Humidity. *Adv. Energy Mater.* **2018**, *8* (3), 1700677.

(5) Fassl, P.; Lami, V.; Bausch, A.; Wang, Z.; Klug, M. T.; Snaith, H. J.; Vaynzof, Y. Fractional Deviations in Precursor Stoichiometry Dictate the Properties, Performance and Stability of Perovskite Photovoltaic Devices. *Energy Environ. Sci.* **2018**, *11* (12), 3380–3391.

(6) Fassl, P.; Ternes, S.; Lami, V.; Zakharko, Y.; Heimfarth, D.; Hopkinson, P. E.; Paulus, F.;
20

Taylor, A. D.; Zaumseil, J.; Vaynzof, Y. Effect of Crystal Grain Orientation on the Rate of Ionic Transport in Perovskite Polycrystalline Thin Films. *ACS Appl. Mater. Interfaces* **2019**, *11* (2), 2490–2499.

(7) Grancini, G.; Srimath Kandada, A. R.; Frost, J. M.; Barker, A. J.; De Bastiani, M.; Gandini, M.; Marras, S.; Lanzani, G.; Walsh, A.; Petrozza, A. Role of Microstructure in the Electron–Hole Interaction of Hybrid Lead Halide Perovskites. *Nat. Photonics* **2015**, *9* (10), 695–701.

(8) Sun, Q.; Fassl, P.; Becker-Koch, D.; Bausch, A.; Rivkin, B.; Bai, S.; Hopkinson, P. E.; Snaith, H. J.; Vaynzof, Y. Role of Microstructure in Oxygen Induced Photodegradation of Methylammonium Lead Triiodide Perovskite Films. *Adv. Energy Mater.* **2017**, *7* (20), 1700977.

(9) Fassl, P.; Zakharko, Y.; Falk, L. M.; Goetz, K. P.; Paulus, F.; Taylor, A. D.; Zaumseil, J.; Vaynzof, Y. Effect of Density of Surface Defects on Photoluminescence Properties in MAPbI 3 Perovskite Films. *J. Mater. Chem. C* **2019**, *7* (18), 5285–5292.

(10) Zhou, H.; Chen, Q.; Li, G.; Luo, S.; Song, T. B.; Duan, H. S.; Hong, Z.; You, J.; Liu, Y.; Yang, Y. Interface Engineering of Highly Efficient Perovskite Solar Cells. *Science* **2014**, *345* (6196), 542–546.

(11) Malinkiewicz, O.; Yella, A.; Lee, Y. H.; Espallargas, G. M.; Graetzel, M.; Nazeeruddin, M. K.; Bolink, H. J. Perovskite Solar Cells Employing Organic Charge-Transport Layers. *Nat. Photonics* **2014**, *8* (2), 128–132.

(12) Zuo, C.; Ding, L. Modified PEDOT Layer Makes a 1.52 V Voc for Perovskite/PCBM Solar Cells. *Adv. Energy Mater.* **2017**, *7* (2), 1601193.

(13) Jeon, N. J.; Noh, J. H.; Kim, Y. C.; Yang, W. S.; Ryu, S.; Seok, S. Il. Solvent Engineering for High-Performance Inorganic–Organic Hybrid Perovskite Solar Cells. *Nat. Mater.* **2014**,

*13* (9), 897–903.

(14) Jeon, N. J.; Lee, H. G.; Kim, Y. C.; Seo, J.; Noh, J. H.; Lee, J.; Seok, S. Il. O -Methoxy Substituents in Spiro-OMeTAD for Efficient Inorganic–Organic Hybrid Perovskite Solar Cells. *J. Am. Chem. Soc.* **2014**, *136* (22), 7837–7840.

(15) Song, Y.; Lv, S.; Liu, X.; Li, X.; Wang, S.; Wei, H.; Li, D.; Xiao, Y.; Meng, Q. Energy Level Tuning of TPB-Based Hole-Transporting Materials for Highly Efficient Perovskite Solar Cells. *Chem. Commun.* **2014**, *50* (96), 15239–15242.

(16) Choi, H.; Paek, S.; Lim, N.; Lee, Y. H.; Nazeeruddin, M. K.; Ko, J. Efficient Perovskite Solar Cells with 13.63 % Efficiency Based on Planar Triphenylamine Hole Conductors. *Chem. - A Eur. J.* **2014**, *20* (35), 10894–10899.

(17) An, Q.; Fassl, P.; Hofstetter, Y. J.; Becker-Koch, D.; Bausch, A.; Hopkinson, P. E.; Vaynzof, Y. High Performance Planar Perovskite Solar Cells by ZnO Electron Transport Layer Engineering. *Nano Energy* **2017**, *39* (April), 400–408.

(18) Liu, Z.; Chang, J.; Lin, Z.; Zhou, L.; Yang, Z.; Chen, D.; Zhang, C.; Liu, S. (Frank); Hao, Y. High-Performance Planar Perovskite Solar Cells Using Low Temperature, Solution-Combustion-Based Nickel Oxide Hole Transporting Layer with Efficiency Exceeding 20%. *Adv. Energy Mater.* **2018**, *8* (19), 1703432.

(19) Song, J.; Zheng, E.; Bian, J.; Wang, X.-F.; Tian, W.; Sanehira, Y.; Miyasaka, T. Low-Temperature $SnO_2$ -Based Electron Selective Contact for Efficient and Stable Perovskite Solar Cells. *J. Mater. Chem. A* **2015**, *3* (20), 10837–10844.

(20) Guo, Q.; Xu, Y.; Xiao, B.; Zhang, B.; Zhou, E.; Wang, F.; Bai, Y.; Hayat, T.; Alsaedi, A.; Tan, Z. Effect of Energy Alignment, Electron Mobility, and Film Morphology of Perylene Diimide Based Polymers as Electron Transport Layer on the Performance of Perovskite

Solar Cells. *ACS Appl. Mater. Interfaces* **2017**, *9* (12), 10983–10991.

(21) Bi, C.; Wang, Q.; Shao, Y.; Yuan, Y.; Xiao, Z.; Huang, J. Non-Wetting Surface-Driven High-Aspect-Ratio Crystalline Grain Growth for Efficient Hybrid Perovskite Solar Cells. *Nat. Commun.* **2015**, *6* (1), 7747.

(22) Wang, Z.-K.; Gong, X.; Li, M.; Hu, Y.; Wang, J.-M.; Ma, H.; Liao, L.-S. Induced Crystallization of Perovskites by a Perylene Underlayer for High-Performance Solar Cells. *ACS Nano* **2016**, *10* (5), 5479–5489.

(23) Arora, N.; Dar, M. I.; Hinderhofer, A.; Pellet, N.; Schreiber, F.; Zakeeruddin, S. M.; Grätzel, M. Perovskite Solar Cells with CuSCN Hole Extraction Layers Yield Stabilized Efficiencies Greater than 20%. *Science* **2017**, *358* (6364), 768–771.

(24) Schulz, P.; Tiepelt, J. O.; Christians, J. A.; Levine, I.; Edri, E.; Sanehira, E. M.; Hodes, G.; Cahen, D.; Kahn, A. High-Work-Function Molybdenum Oxide Hole Extraction Contacts in Hybrid Organic-Inorganic Perovskite Solar Cells. *ACS Appl. Mater. Interfaces* **2016**, *8* (46), 31491–31499.

(25) Sanehira, E. M.; Tremolet de Villers, B. J.; Schulz, P.; Reese, M. O.; Ferrere, S.; Zhu, K.; Lin, L. Y.; Berry, J. J.; Luther, J. M. Influence of Electrode Interfaces on the Stability of Perovskite Solar Cells: Reduced Degradation Using MoO x /Al for Hole Collection. *ACS Energy Lett.* **2016**, *1* (1), 38–45.

(26) Zhao, Y.; Nardes, A. M.; Zhu, K. Effective Hole Extraction Using MoOx-Al Contact in Perovskite CH3NH3PbI3 Solar Cells. *Appl. Phys. Lett.* **2014**, *104* (21).

(27) Chang, C.-Y.; Chang, Y.-C.; Huang, W.-K.; Lee, K.-T.; Cho, A.-C.; Hsu, C.-C. Enhanced Performance and Stability of Semitransparent Perovskite Solar Cells Using Solution-Processed Thiol-Functionalized Cationic Surfactant as Cathode Buffer Layer. *Chem. Mater.*

**2015**, *27* (20), 7119–7127.

(28) Peng, S.; Miao, J.; Murtaza, I.; Zhao, L.; Hu, Z.; Liu, M.; Yang, T.; Liang, Y.; Meng, H.; Huang, W. An Efficient and Thickness Insensitive Cathode Interface Material for High Performance Inverted Perovskite Solar Cells with 17.27% Efficiency. *J. Mater. Chem. C* **2017**, *5* (24), 5949–5955.

(29) Zhang, H.; Azimi, H.; Hou, Y.; Ameri, T.; Przybilla, T.; Spiecker, E.; Kraft, M.; Scherf, U.; Brabec, C. J. Improved High-Efficiency Perovskite Planar Heterojunction Solar Cells via Incorporation of a Polyelectrolyte Interlayer. *Chem. Mater.* **2014**, *26* (18), 5190–5193.

(30) Docampo, P.; Ball, J. M.; Darwich, M.; Eperon, G. E.; Snaith, H. J. Efficient Organometal Trihalide Perovskite Planar-Heterojunction Solar Cells on Flexible Polymer Substrates. *Nat. Commun.* **2013**, *4* (1), 2761.

(31) Rivkin, B.; Fassl, P.; Sun, Q.; Taylor, A. D.; Chen, Z.; Vaynzof, Y. Effect of Ion Migration-Induced Electrode Degradation on the Operational Stability of Perovskite Solar Cells. *ACS Omega* **2018**, *3* (8), 10042–10047.

(32) Chen, G.; Zhang, F.; Liu, M.; Song, J.; Lian, J.; Zeng, P.; Yip, H.-L.; Yang, W.; Zhang, B.; Cao, Y. Fabrication of High-Performance and Low-Hysteresis Lead Halide Perovskite Solar Cells by Utilizing a Versatile Alcohol-Soluble Bispyridinium Salt as an Efficient Cathode Modifier. *J. Mater. Chem. A* **2017**, *5* (34), 17943–17953.

(33) Wang, Q.; Shao, Y.; Dong, Q.; Xiao, Z.; Yuan, Y.; Huang, J. Large Fill-Factor Bilayer Iodine Perovskite Solar Cells Fabricated by a Low-Temperature Solution-Process. *Energy Environ. Sci.* **2014**, *7* (7), 2359–2365.

(34) Liu, D.; Wang, Q.; Traverse, C. J.; Yang, C.; Young, M.; Kuttipillai, P. S.; Lunt, S. Y.; Hamann, T. W.; Lunt, R. R. Impact of Ultrathin C 60 on Perovskite Photovoltaic Devices.

*ACS Nano* **2018**, *12* (1), 876–883.

(35) Zhou, L.; Chang, J.; Liu, Z.; Sun, X.; Lin, Z.; Chen, D.; Zhang, C.; Zhang, J.; Hao, Y. Enhanced Planar Perovskite Solar Cell Efficiency and Stability Using a Perovskite/PCBM Heterojunction Formed in One Step. *Nanoscale* **2018**, *10* (6), 3053–3059.

(36) Zhang, S.; Stolterfoht, M.; Armin, A.; Lin, Q.; Zu, F.; Sobus, J.; Jin, H.; Koch, N.; Meredith, P.; Burn, P. L.; Neher, D. Interface Engineering of Solution-Processed Hybrid Organohalide Perovskite Solar Cells. *ACS Appl. Mater. Interfaces* **2018**, *10* (25), 21681–21687.

(37) Zhao, L.; Luo, D.; Wu, J.; Hu, Q.; Zhang, W.; Chen, K.; Liu, T.; Liu, Y.; Zhang, Y.; Liu, F.; Russell, T. P.; Snaith, H. J.; Zhu, R.; Gong, Q. High-Performance Inverted Planar Heterojunction Perovskite Solar Cells Based on Lead Acetate Precursor with Efficiency Exceeding 18%. *Adv. Funct. Mater.* **2016**, *26* (20), 3508–3514.

(38) Kim, S.-M.; Kim, C.-H.; Kim, Y.; Kim, N.; Lee, W.-J.; Lee, E.-H.; Kim, D.; Park, S.; Lee, K.; Rivnay, J.; Yoon, M.-H. Influence of PEDOT:PSS Crystallinity and Composition on Electrochemical Transistor Performance and Long-Term Stability. *Nat. Commun.* **2018**, *9* (1), 3858.

(39) An, Q.; Sun, Q.; Weu, A.; Becker-Koch, D.; Paulus, F.; Arndt, S.; Stuck, F.; Hashmi, A. S. K.; Tessler, N.; Vaynzof, Y. Enhancing the Open-Circuit Voltage of Perovskite Solar Cells by up to 120 MV Using Π-Extended Phosphoniumfluorene Electrolytes as Hole Blocking Layers. *Adv. Energy Mater.* **2019**, *9* (33), 1901257.

(40) Intorp, S. N.; Kushida, S.; Dmitrieva, E.; Popov, A. A.; Rominger, F.; Freudenberg, J.; Hinkel, F.; Bunz, U. H. F. True Blue Through Oxidation—A Thiaazulenic Heterophenoquinone as Electrochrome. *Chem. - A Eur. J.* **2019**, *25* (21), 5412–5415.

(41) Wudarczyk, J.; Papamokos, G.; Margaritis, V.; Schollmeyer, D.; Hinkel, F.; Baumgarten,

M.; Floudas, G.; Müllen, K. Hexasubstituted Benzenes with Ultrastrong Dipole Moments. *Angew. Chemie - Int. Ed.* **2016**, *55* (9), 3220–3223.

(42) Giovambattista, N.; Debenedetti, P. G.; Rossky, P. J. Effect of Surface Polarity on Water Contact Angle and Interfacial Hydration Structure. *J. Phys. Chem. B* **2007**, *111* (32), 9581–9587.

(43) Fang, R.; Wu, S.; Chen, W.; Liu, Z.; Zhang, S.; Chen, R.; Yue, Y.; Deng, L.; Cheng, Y.-B.; Han, L.; Chen, W. [6,6]-Phenyl-C 61 -Butyric Acid Methyl Ester/Cerium Oxide Bilayer Structure as Efficient and Stable Electron Transport Layer for Inverted Perovskite Solar Cells. *ACS Nano* **2018**, *12* (3), 2403–2414.

(44) Tessler, N.; Vaynzof, Y. Preventing Hysteresis in Perovskite Solar Cells by Undoped Charge Blocking Layers. *ACS Appl. Energy Mater.* **2018**, *1* (2), 676–683.

(45) Schulz, P. Interface Design for Metal Halide Perovskite Solar Cells. *ACS Energy Lett.* **2018**, *3* (6), 1287–1293.

(46) Sarritzu, V.; Sestu, N.; Marongiu, D.; Chang, X.; Masi, S.; Rizzo, A.; Colella, S.; Quochi, F.; Saba, M.; Mura, A.; Bongiovanni, G. Optical Determination of Shockley-Read-Hall and Interface Recombination Currents in Hybrid Perovskites. *Sci. Rep.* **2017**, *7* (1), 44629.

(47) Zarazua, I.; Han, G.; Boix, P. P.; Mhaisalkar, S.; Fabregat-Santiago, F.; Mora-Seró, I.; Bisquert, J.; Garcia-Belmonte, G. Surface Recombination and Collection Efficiency in Perovskite Solar Cells from Impedance Analysis. *J. Phys. Chem. Lett.* **2016**, *7* (24), 5105–5113.

(48) Nimens, W. J.; Ogle, J.; Caruso, A.; Jonely, M.; Simon, C.; Smilgies, D.; Noriega, R.; Scarpulla, M.; Whittaker-Brooks, L. Morphology and Optoelectronic Variations Underlying the Nature of the Electron Transport Layer in Perovskite Solar Cells. *ACS Appl.*

*Energy Mater.* **2018**, *1* (2), 602–615.

(49) Falk, L. M.; Goetz, K. P.; Lami, V.; An, Q.; Fassl, P.; Herkel, J.; Thome, F.; Taylor, A. D.; Paulus, F.; Vaynzof, Y. Effect of Precursor Stoichiometry on the Performance and Stability of MAPbBr 3 Photovoltaic Devices. *Energy Technol.* **2019**, *1900737*, ente.201900737.

(50) Stolterfoht, M.; Wolff, C. M.; Amir, Y.; Paulke, A.; Perdigón-Toro, L.; Caprioglio, P.; Neher, D. Approaching the Fill Factor Shockley-Queisser Limit in Stable, Dopant-Free Triple Cation Perovskite Solar Cells. *Energy Environ. Sci.* **2017**, *10* (6), 1530–1539.

(51) Bu, T.; Liu, X.; Zhou, Y.; Yi, J.; Huang, X.; Luo, L.; Xiao, J.; Ku, Z.; Peng, Y.; Huang, F.; Cheng, Y. B.; Zhong, J. A Novel Quadruple-Cation Absorber for Universal Hysteresis Elimination for High Efficiency and Stable Perovskite Solar Cells. *Energy Environ. Sci.* **2017**, *10* (12), 2509–2515.

(52) Belisle, R. A.; Jain, P.; Prasanna, R.; Leijtens, T.; McGehee, M. D. Minimal Effect of the Hole-Transport Material Ionization Potential on the Open-Circuit Voltage of Perovskite Solar Cells. *ACS Energy Lett.* **2016**, *1* (3), 556–560.

(53) Magen, O.; Tessler, N. Charge Blocking Layers in Thin-Film/Amorphous Photovoltaics. *J. Appl. Phys.* **2016**, *120* (19), 194502.

(54) Tessler, N. Adding 0.2 v to the Open Circuit Voltage of Organic Solar Cells by Enhancing the Built-in Potential. *J. Appl. Phys.* **2015**, *118* (21), 215501.

(55) Luo, D.; Yang, W.; Wang, Z.; Sadhanala, A.; Hu, Q.; Su, R.; Shivanna, R.; Trindade, G. F.; Watts, J. F.; Xu, Z.; Liu, T.; Chen, K.; Ye, F.; Wu, P.; Zhao, L.; Wu, J.; Tu, Y.; Zhang, Y.; Yang, X.; Zhang, W.; Friend, R. H.; Gong, Q.; Snaith, H. J.; Zhu, R. Enhanced Photovoltage for Inverted Planar Heterojunction Perovskite Solar Cells. *Science* **2018**, *360* (6396), 1442–1446.

(56) Wang, Q.; Bi, C.; Huang, J. Doped Hole Transport Layer for Efficiency Enhancement in Planar Heterojunction Organolead Trihalide Perovskite Solar Cells. *Nano Energy* **2015**, *15*, 275–280.

(57) Zhao, J.; Zheng, X.; Deng, Y.; Li, T.; Shao, Y.; Gruverman, A.; Shield, J.; Huang, J. Is Cu a Stable Electrode Material in Hybrid Perovskite Solar Cells for a 30-Year Lifetime? *Energy Environ. Sci.* **2016**, *9* (12), 3650–3656.

(58) Chen, W.; Zhou, Y.; Wang, L.; Wu, Y.; Tu, B.; Yu, B.; Liu, F.; Tam, H. W.; Wang, G.; Djurišić, A. B.; Huang, L.; He, Z. Molecule-Doped Nickel Oxide: Verified Charge Transfer and Planar Inverted Mixed Cation Perovskite Solar Cell. *Adv. Mater.* **2018**, *30* (20), 1–9.

(59) Zheng, J.; Hu, L.; Yun, J. S.; Zhang, M.; Lau, C. F. J.; Bing, J.; Deng, X.; Ma, Q.; Cho, Y.; Fu, W.; Chen, C.; Green, M. A.; Huang, S.; Ho-Baillie, A. W. Y. Solution-Processed, Silver-Doped NiO x as Hole Transporting Layer for High-Efficiency Inverted Perovskite Solar Cells. *ACS Appl. Energy Mater.* **2018**, *1* (2), 561–570.

(60) Park, J. H.; Seo, J.; Park, S.; Shin, S. S.; Kim, Y. C.; Jeon, N. J.; Shin, H.-W.; Ahn, T. K.; Noh, J. H.; Yoon, S. C.; Hwang, C. S.; Seok, S. Il. Efficient CH 3 NH 3 PbI 3 Perovskite Solar Cells Employing Nanostructured P-Type NiO Electrode Formed by a Pulsed Laser Deposition. *Adv. Mater.* **2015**, *27* (27), 4013–4019.